\documentclass[letterpaper]{article} % DO NOT CHANGE THIS
\usepackage{aaai25}  % DO NOT CHANGE THIS
\usepackage{times}  % DO NOT CHANGE THIS
\usepackage{helvet}  % DO NOT CHANGE THIS
\usepackage{courier}  % DO NOT CHANGE THIS
\usepackage[hyphens]{url}  % DO NOT CHANGE THIS
\usepackage{graphicx} % DO NOT CHANGE THIS
\usepackage{natbib}  % DO NOT CHANGE THIS AND DO NOT ADD ANY OPTIONS TO IT
\usepackage{caption} % DO NOT CHANGE THIS AND DO NOT ADD ANY OPTIONS TO IT
\usepackage{algorithm}

\usepackage[table,xcdraw]{xcolor}

\usepackage{algorithm}

\usepackage{url}
\usepackage{booktabs}
\usepackage{amsfonts}
\usepackage{nicefrac}
\usepackage{microtype}
\usepackage{xcolor}
\usepackage{commath}

\usepackage{graphicx}
\usepackage{amsmath}
\usepackage{amsthm}
\usepackage{algpseudocode}
\usepackage{appendix}
\usepackage{multirow}

\newtheorem{definition}{Definition}
\newtheorem{theorem}{Theorem}
\newtheorem{lemma}{Lemma}

\usepackage{paralist}

\usepackage{newfloat}
\usepackage{listings}
\DeclareCaptionStyle{ruled}{labelfont=normalfont,labelsep=colon,strut=off} % DO NOT CHANGE THIS
\floatstyle{ruled}
\newfloat{listing}{tb}{lst}{}
\floatname{listing}{Listing}
\nocopyright 

\title{Scalable Quantum-Inspired Optimization through Dynamic Qubit Compression}
\author{
    Co Tran\equalcontrib\textsuperscript{\rm 1}-
    Quoc-Bao Tran\equalcontrib\textsuperscript{\rm 2},
    Hy Truong Son\textsuperscript{\rm 3},
    Thang N Dinh\textsuperscript{\rm 2}\thanks{Corresponding author: tndinh@vcu.edu}
}
\affiliations{
    \textsuperscript{\rm 1}University of Texas at Austin\\
    \textsuperscript{\rm 2}Virginia Commonwealth University\\
    \textsuperscript{\rm 3}University of Alabama at Birmingham\\
    co.quoc.tran.2@gmail.com, tranq3@vcu.edu, thy@uab.edu, tndinh@vcu.edu
}

\usepackage{bibentry}
\begin{document}

\maketitle

\begin{abstract}

Hard combinatorial optimization problems, often mapped to Ising models, promise potential solutions with quantum advantage but are constrained by limited qubit counts in near-term devices. We present an innovative quantum-inspired framework that dynamically compresses large Ising models to fit available quantum hardware of different sizes.  Thus, we aim to bridge the gap between large-scale optimization and current hardware capabilities. Our method leverages a physics-inspired GNN architecture to capture complex interactions in Ising models and accurately predict alignments among neighboring spins (aka qubits) at ground states. By progressively merging such aligned spins, we can reduce the model size while preserving the underlying optimization structure. It also provides a natural trade-off between the solution quality and size reduction, meeting different hardware constraints of quantum computing devices. Extensive numerical studies on  Ising instances of diverse topologies show that our method can reduce instance size at multiple levels with virtually no losses in solution quality on the latest D-wave quantum annealers.
\end{abstract}

\section{Introduction} \label{sec:intro}
Combinatorial optimization problems are ubiquitous in various domains, including portfolio optimization \cite{mugel2021hybrid,grozea2021optimising}, car manufacturing scheduling \cite{yarkoni2021multi}, and RNA folding \cite{fox2021mrna,fox2022rna}. These problems often involve finding the optimal solution among a vast number of possibilities, making them computationally challenging. Many of these problems can be mapped to Ising models \cite{lucas2014ising}, which encode the optimization objective in terms of interacting spins. However, a significant number of these problems fall into the NP-hard complexity class \cite{gary1979computers}, meaning they are intractable for classical computers as the problem size grows. This intractability has motivated the exploration of alternative computing paradigms, such as quantum annealing \cite{kadowaki1998quantum,zhou2022noise}, which leverages quantum mechanics to potentially solve these problems more efficiently than classical methods.

Recent years have witnessed remarkable advances in quantum computing, bringing us closer to the realm of "quantum supremacy" \cite{Preskill_2018}, where quantum processors solve problems intractable for classical computers. Notable milestones include Google's Sycamore processor demonstrating supremacy in a sampling task \cite{arute2019quantum} and China's Jiuzhang photonic quantum computer achieving quantum advantage for Gaussian boson sampling \cite{zhong2020quantum}. Beyond these proof-of-concept demonstrations, there have been efforts to showcase quantum utility on more practical problems. Quantum annealing, implemented in D-Wave's systems \cite{Boothby2020}, has demonstrated quantum advantages for certain types of problems \cite{king2021scaling, Tasseff2022, tasseff2024emerging, king2024computational}. Gate-based algorithms such as the Quantum Approximate Optimization Algorithm (QAOA) \cite{bauza2024scaling} and Variational Quantum Eigensolver (VQE) \cite{peruzzo2014variational} offer alternative routes for tackling optimization challenges. Additionally, quantum-inspired specialized hardware, including optical Ising machines \cite{honjo2021100}, digital annealers, and FPGA-based solvers \cite{patel2020ising}, provide complementary approaches to address complex optimization problems.

Despite the rapid progress in quantum computing, qubit count remains a significant limiting factor for solving practical optimization problems. Current state-of-the-art quantum annealers, such as D-Wave's Advantage platform, offer over 5000 qubits \cite{Boothby2020}. However, many real-world applications require even more qubits. For example, performing MIMO channel decoding with a 60Tx60R setup on a 64-QAM configuration would necessitate about 11,000 physical qubits \cite{tabi2021evaluation}, exceeding the capabilities of existing hardware. The challenge is further compounded by limited qubit connectivity, which necessitates complex minor embedding techniques \cite{choi2008minor,choi2011minor}, significantly increasing the number of physical qubits required. These hardware constraints substantially limit the size and complexity of problems that can be directly solved on quantum processors with a clear advantage over classical methods.

While waiting for quantum hardware advances, a parallel challenge emerges: efficiently reducing Ising models to fit limited qubit capacities. Current reduction techniques, ranging from classical roof duality \cite{hammer1984roof,boros2002pseudo} and extended roof duality \cite{rother2007optimizing} to recent graph-based approaches \cite{thai2022fasthare}, show promise but face significant limitations. These methods, constrained by the need for an exact reduction, can compress only a fraction of problem instances—with the best-performing algorithms reducing less than half of tested cases, achieving an average compression of about 20\% among instances \cite{thai2022fasthare}. This variability in effectiveness highlights a notable gap: the absence of a universal, tunable reduction method to compress Ising models to arbitrary sizes. Such flexibility could accommodate diverse quantum hardware constraints and potentially enhance the applicability of quantum annealing across a broader spectrum of real-world optimization problems.

We present a novel framework that leverages Graph Neural Networks (GNNs) to dynamically compress large Ising models for quantum annealing. Our approach automates the discovery of combinatorial rules for qubit reduction by training a GNN to predict ground-state qubit alignments and identify optimal contraction candidates. This data-driven method enables progressive spin merging while preserving solution integrity, offering a tunable trade-off between compression ratio and solution quality. Unlike previous compression techniques that relied on manual heuristics \cite{thai2022fasthare}, our GNN-based approach captures complex patterns in both local and global spin interactions, identifying compressible qubit groups that elude detection by conventional methods. In contrast to GNN-based methods to directly solve Ising models and combinatorial optimization problems \cite{Dai2017LC, Li2018COG, Gasse2019, Joshi2019AnEG, Schuetz2021PIGNN, Schuetz2022}, these methods face challenges, including sensitivity to graph structure and connectivity, and poor performance on sparse graphs \cite{Panfeng2021}. Moreover, by acting as a preprocessing phase for quantum computing, our approaches preserve the potential for a quantum advantage.

\paragraph{Our contributions.}We summarize below our contributions
\begin{itemize}
\item We present GRANITE, a GNN-based framework that dynamically compresses large Ising models, automating the discovery of qubit reduction rules. This method efficiently predicts ground state alignments and identifies optimal contractions, offering tunable trade-offs between model size and solution quality, and accommodating diverse quantum hardware constraints. The compression preserves the Ising structure and can work with any quantum technology that solves Ising models, including both quantum annealers and gate-based quantum computers via Quantum Approximate Optimization Algorithm (QAOA).
\item Our extensive testing reveals substantial multi-level size reductions across various Ising topologies while preserving solution accuracy.
\item By significantly reducing qubit requirements, our approach expands the scope of tractable problems for current quantum annealers, potentially accelerating practical quantum advantage in optimization tasks. This work provides a powerful tool for exploring the quantum-classical computational boundary, addressing a critical challenge in near-term quantum computing.
\end{itemize}
This work addresses directly the qubit limitation challenge, offering a powerful, flexible tool for researchers and practitioners in quantum optimization.

\begin{figure*}[h]
\centering
  \includegraphics[width=0.7\textwidth]{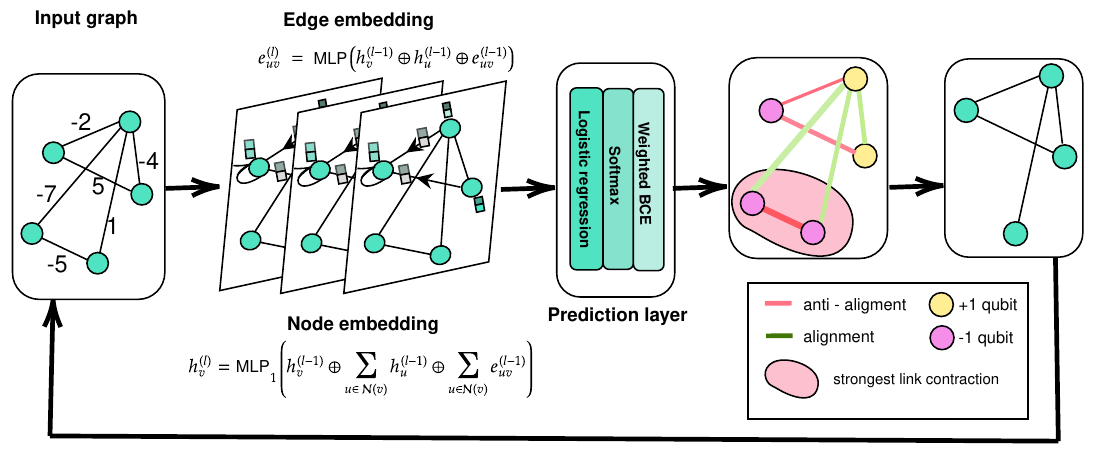}
  \caption{\textbf{GRANITE: Graph Neural Ising Transformer for Efficient Quantum Optimization.} The model comprises three key components: \textbf{a,} A GNN that learns edge ($e_{uv}$) and node ($h_v$) representations, capturing the Ising model's structure and interactions. \textbf{b,} A prediction layer using logistic regression with softmax to calculate weighted binary cross-entropy, assigning confidence scores to potential actions. \textbf{c,} A link contraction process that executes the highest-confidence merge or flip-merge operation. During inference, the contracted graph is iteratively fed back into GRANITE until the desired reduction ratio is achieved, enabling the transformation of large-scale Ising problems into quantum-compatible formats.}
  \label{fig:granite}
\end{figure*}

\section{Related work} \label{sec:related}

\paragraph{Ising Models and Combinatorial Optimization.}
Ising models, which naturally lend themselves to graph representations, have been a focal point in statistical physics and combinatorial optimization \cite{Carleo2019, Akinori2023}. These models are particularly challenging due to their NP-hardness and have applications across various domains, including computer science and machine learning.
\paragraph{Traditional Approaches to Solving Ising Models.}
Solving computationally difficult Ising models has traditionally relied on heuristic algorithms and physics-inspired techniques. Simulated Annealing (SA) \cite{kirkpatrick1983optimization} has been a cornerstone approach. More recently, Ising machines based on algorithms such as SimCIM (simulated coherent Ising machine) \cite{tiunov2019annealing, king2018emulating} and simulated bifurcation (SB) \cite{goto2021high, oshiyama2022benchmark} have shown impressive results in finding ground states of Ising models.
\paragraph{Machine Learning Approaches for Ising Models.}
Machine learning techniques have been increasingly applied to Ising problems. Variational Autoregressive Networks (VANs) \cite{wu2019solving} and Variational Classical Annealing (VCA) \cite{hibat2021variational, hibat2020recurrent} have shown promise. VCA, in particular, outperforms traditional SA but faces scalability issues, being limited to problems with up to 32 spin variables in challenging scenarios like the Wishart Planted Ensemble (WPE) \cite{hamze2020wishart}. Reinforcement learning approaches \cite{angelini2023modern, panchenko2013sherrington} offer an alternative by directly optimizing for ground state configurations.
\paragraph{Graph Neural Networks for Ising Models.}
Graph Neural Networks (GNNs) have emerged as a promising approach for solving Ising models and related combinatorial optimization problems \cite{Dai2017LC, Li2018COG, Gasse2019, Joshi2019AnEG, Schuetz2021PIGNN, Schuetz2022}. Some GNN-based methods have demonstrated the ability to handle large-scale instances with millions of variables \cite{Schuetz2021PIGNN}. However, these approaches face challenges, including sensitivity to graph structure and connectivity, and poor performance on sparse graphs \cite{Panfeng2021}. Recent works have questioned the effectiveness of GNNs compared to classical heuristic algorithms for certain problems \cite{Boettcher2022, Angelini2022}.
Our proposed approach, unlike existing GNN-based methods that attempt to solve Ising models directly, focuses on using GNNs to compress Ising models. This novel perspective aims to address the scalability issues faced by current methods while maintaining the ability to capture complex interactions in the Ising system.\section{Background} \label{sec:background}
\subsection{Ising Models and NP-hard Problems}
Ising models, originally developed in statistical physics, have become a powerful framework for representing combinatorial optimization problems \cite{lucas2014ising}. An Ising model consists of binary variables (spins) $s_i \in {-1, +1}$, with energy given by the Hamiltonian:
\begin{equation}
\label{eq:Ising}
H = -\sum_{i,j} J_{ij} s_i s_j - \sum_i h_i s_i
\end{equation}
where $J_{ij}$ represents the coupling strength between spins $i$ and $j$, and $h_i$ is the external field acting on spin $i$.
Many NP-hard problems can be reformulated as Quadratic Unconstrained Binary Optimization (QUBO) problems \cite{glover2018tutorial}, which are closely related to Ising models. The QUBO formulation uses binary variables $x_i \in {0, 1}$:
\begin{equation}
\min_{x} \sum_{i,j} Q_{ij} x_i x_j + \sum_i c_i x_i
\end{equation}
QUBO can be mapped to Ising models through the transformation: $s_i = 2x_i - 1$.
NP-hard problems are characterized by exponential time complexity for exact classical algorithms \cite{gary1979computers}. This intractability for large instances has motivated the exploration of quantum annealing as an alternative computing paradigm \cite{kadowaki1998quantum}. By mapping NP-hard problems to Ising models, researchers aim to leverage quantum effects to explore complex solution spaces more efficiently, potentially overcoming limitations of classical computing such as exponential state space growth and local optima traps.

\subsection{Quantum Annealing}

Quantum annealing (QA) is a metaheuristic for solving optimization problems that leverages quantum mechanical effects such as tunneling and superposition \cite{Kadowaki_1998, Farhi2000, Santoro2006}. It is particularly well-suited for solving problems formulated as Ising models or Quadratic Unconstrained Binary Optimization (QUBO) problems, as discussed in the previous section.

The quantum annealing process is based on adiabatic quantum computation. The system is initialized in the ground state of an easily prepared Hamiltonian, typically the transverse field Hamiltonian, and then evolves according to:

\begin{equation}
    H_\text{system}(s) = - \frac{A(s)}{2} \Big( \sum_i^n \sigma^x_i \Big) + \frac{B(s)}{2} \Big( H_\text{problem} \Big)
\end{equation}
where $s \in [0,1]$ is the anneal fraction, $\sigma^x_i$ is the Pauli x-matrix for the i-th qubit, and $H_\text{problem}$ is the problem Hamiltonian, equivalent to the Ising model Hamiltonian from eq.~\eqref{eq:Ising}. $A(s)$ and $B(s)$ define the anneal schedule, with $A(s)$ decreasing and $B(s)$ increasing as $s$ goes from 0 to 1 \cite{Hauke_2020}.

Quantum annealing harnesses quantum effects like superposition to tackle optimization problems more efficiently than classical methods. By exploring multiple states simultaneously and "tunneling" through energy barriers, it can potentially find better solutions that classical algorithms might miss.  Recent studies have demonstrated a quantum advantage for certain problem classes \cite{harris2018phase, King_2022, king2024computational}, highlighting the potential of QA as a powerful tool for solving Ising model-based optimization problems.

Despite the promise of quantum annealing, current systems face several challenges including the key challenge of limited qubit counts. The latest quantum annealer from D-Wave  features fewer than 6000 flux qubits arranged, in a Pegasus topology \cite{Boothby2020}.
As quantum annealing technology continues to advance, it promises to tackle increasingly complex optimization problems that are intractable for classical computers.\section{Method} \label{sec:method}

\subsection{Qubit Alignment at Ground States}
Ising models, fundamental in statistical physics and optimization, can be elegantly represented as graphs. This representation not only captures the model's structure but also enables powerful reduction techniques.

\paragraph{Graph representation.} Consider an Ising Hamiltonian $(h, J)$ over spins $s_1$ to $s_n$. We construct an undirected weighted graph $G_H = (V, E)$ where $V = \{0, 1, \ldots, n\}$ and $E = \{(i, j) \mid J_{ij} \neq 0\} \cup \{(0, i) \mid h_i \neq 0\}$ (Fig. 1a). The auxiliary vertex 0 represents linear biases, unifying the treatment of linear and quadratic interactions.
Edge weights encapsulate both interaction types:
\begin{equation}
w(i, j) = \begin{cases}
    J_{ij} & \text{if } i \neq j \\
    h_j & \text{if } i = 0
\end{cases}
\end{equation}
Our choice for representation simplifies the analysis by treating all interactions uniformly as edge weights.

\paragraph{Qubits alignment.} Ground states, configurations with minimum energy, reveal crucial structural information. We classify each edge $(i, j)$ based on the behavior of connected spins across all ground states:
\begin{itemize}
\item \textbf{Alignment}: $s_i$ and $s_j$ always have the same value.
\item \textbf{Anti-alignment}: $s_i$ and $s_j$ always have different values.
\item \textbf{Neutral}: $s_i$ and $s_j$ alignment varies among ground states.
\end{itemize}
This classification enables targeted graph reductions that preserve ground-state properties.
For an aligned edge $(i, j)$, we have $s_i = s_j$ in all ground-states, thus, we can replace $s_i = s_j$ and remove $s_i$ from the Ising. Equivalently, we perform a merge operation on the graph. This operation combines two nodes that always have the same value in ground states, effectively reducing the number of variables in our system. The merge operation removes one node and redirects its connections to the remaining node. Formally, we define the merge operation $M(i, j)$ as:
\begin{align*}
M(i, j): &V' = V \setminus \{j\}, \\
&E' = \{(k, l) \in E \mid k, l \neq j\} \cup \{(i, k) \mid (j, k) \in E\}\\
&w(i, k) = w(i ,k) + w(j, k)
\end{align*}

Anti-aligned edge $(i, j)$ means $s_i = -s_j$ in all ground-states, thus, we can replace $s_i$ with $-s_j$ and remove $s_i$, completely. In the graph, the two nodes $i$ and $j$ undergo a two-step flip-merge operation. This operation first flips the sign of all interactions involving one node (to account for the constant difference in spin values), then merges the nodes. 

The flip-merge operation $FM(i, j)$ consists of:
\begin{enumerate}
\item \textit{Flip}: Negate weights of edges incident to $j$:\\ $w(j, k) = -w(j, k)\ \forall (k, j) \in E$.
\item \textit{Merge}: Apply $M(i, j)$ as defined above.
\end{enumerate}

Finally, neutral edges allow either merge or flip-merge, offering flexibility in reduction strategy.

These operations can significantly simplify Ising Hamiltonians while maintaining their essential properties. Iterative application potentially reduces problem complexity, guiding the development of efficient solution methods or revealing underlying system structure.

\paragraph{Hardness of Predicting Qubit Alignment.} Predicting the alignment of qubits in Ising Hamiltonian is, however, intractable.
\begin{theorem}
\label{theo:co-np-hard}
The problem of classifying a single edge in an Ising model as alignment or non-alignment is Co-NP-hard. Consequently, there is no polynomial-time algorithm for this problem unless P = NP.
\end{theorem}

The proof is done by a polynomial-time reduction from a Co-NP-complete problem of determining whether all truth assignments of a 3-SAT formula satisfy $x_i = x_j$ (or alternatively, $x_i \neq x_j$) for some pair of variables $x_i$ and $x_j$. Given an instance of this problem with a 3-SAT formula $\phi(x_1, \ldots, x_n)$, we construct an Ising model $H$ by mapping each variable $x_k$ to a spin $s_k$.  The question of whether all truth assignments of $\phi$ satisfy $x_i = x_j$ is then equivalent to asking whether $s_i$ and $s_j$ are aligned in all ground states of $H$. Thus, if we could efficiently classify the edge $(i,j)$ as alignment or non-alignment, we could solve the original Co-NP-complete problem in polynomial time. This reduction, combined with the fact that the problem is clearly in Co-NP (a counterexample of non-alignment can be verified in polynomial time), establishes that edge classification is Co-NP-complete. A complete proof with detailed construction of the Hamiltonian and analysis is provided in the Supplementary Information.

The Co-NP-completeness of edge classification in Ising models underscores its computational intractability, with exact solutions requiring exhaustive examination of all spin configurations to identify the complete set of ground states. While this approach remains viable for small instances, it becomes infeasible as system size grows, limiting its applicability in practical scenarios.
This computational barrier motivates the exploration of alternative strategies, particularly in the realm of machine learning.

\subsection{\textbf{G}raph \textbf{N}eural \textbf{I}sing \textbf{T}ransformer for \textbf{E}fficient Quantum Optimization (GRANITE)
}
\label{subsection:link}
Graph Neural Networks (GNNs) emerge as a promising candidate, offering a unique ability to capture the intricate spatial relationships and interactions inherent in Ising models. By leveraging the graph structure of the Ising Hamiltonian, GNNs can potentially learn to approximate edge classifications without explicit enumeration of ground states, opening avenues for scalable analysis of larger systems. This approach not only promises computational efficiency but also the potential to uncover hidden patterns and heuristics in edge behavior across diverse Ising instances, potentially leading to new insights into the structure of complex spin systems.

The \textbf{G}raph \textbf{N}eural \textbf{I}sing \textbf{T}ransformer for \textbf{E}fficient Quantum Optimization (GRANITE) leverages Graph Neural Networks (GNNs) to navigate the complex landscape of Ising model reduction (Fig.~\ref{fig:granite}). GRANITE iteratively predicts optimal graph contraction operations—merge or flip-merge—for each edge in the Ising model's graph representation. In each iteration, the GNN processes the current graph structure, learning edge and node representations that capture local and global spin interactions. These representations feed into a prediction layer, which assigns confidence scores to potential merge and flip-merge operations for each edge. The edge with the highest confidence score is selected, and its associated operation is performed, reducing the graph by one node. This process repeats until the Ising model is sufficiently small to be handled by quantum hardware, effectively bridging the gap between large-scale classical problems and limited-size quantum processors. The GRANITE workflow can be summarized as follows:

\begin{algorithm}
\caption{GRANITE - Graph Neural Ising Transformer}
\begin{algorithmic}[1]
\Require
    \State Ising model graph $G=(V,E)$
    \State Desired reduction ratio $\alpha$
    \State Number of GNN layers $L$
\Ensure Reduced Ising model graph $G'$
\State Initialize the GNN model with specified hyperparameters
\While{$\text{size}(G) > \alpha \times (\text{initial size of } G)$}
    \For{$\ell = 1$ to $L$}

            \State Compute node representation $h_v^{(\ell)}, \ \forall v \in V.$

            \State Compute edge representation $e_{uv}^{(\ell)}, \ \forall (u, v) \in E.$

    \EndFor

    \For{each edge $(u,v) \in E$}
        \State $z_{uv} = h_u^{(L)} \oplus h_v^{(L)} \oplus e_{uv}^{(L)}$
        \State $\hat{y}_{uv} = \sigma(\langle w, z_{uv} \rangle)$
        \State $C_{uv} = -(\hat{y}_{uv}\log(\hat{y}_{uv}) + (1-\hat{y}_{uv})\log(1-\hat{y}_{uv}))$
    \EndFor

    \State $(\hat u,\hat  v) = \arg\max_{(u,v) \in E} C_{uv}$
    \If{$\hat{y}_{\hat u \hat v} < 0.5$} \Comment{Regular merge}
        \State $G = \text{Merge}(G, \hat u, \hat v)$
    \Else \Comment{Flip then merge}
        \State $G = \text{Flip}(G, \hat u)$
        \State $G = \text{Merge}(G, \hat u, \hat v)$
    \EndIf
\EndWhile
\State \Return $G$
\end{algorithmic}
\end{algorithm}

The key advantage of leveraging GNNs in this process is their ability to learn complex graph structures and capture both local and global information, enabling accurate identification of non-separable qubit groups. By exploiting the representational power of GNNs, our approach aims to improve upon the greedy merging strategies employed in previous methods to lead to more effective Qubits reduction and facilitate the solution of larger
optimization problems on quantum annealing hardware.

\paragraph{GNN Model.} We start with the representation of Ising Model with Hamiltonian graph $G_H = (V, E)$ where $V$ is the set of nodes and $E \subseteq V \times V$ is the set of edges. We construct the initial node and edge features are defined as:
$$\mathbf{H}^{(0)} = \{h_v^{(0)} \in \mathbb{R}^{d_h} \mid v \in V\},$$
$$\mathbf{E}^{(0)} = \{e_{uv}^{(0)} \in \mathbb{R}^{d_e} \mid (u, v) \in E\},$$
respectively, in which $d_h$ and $d_e$ are the corresponding numbers of input node and edge features. For each node $v$, $h_v^{(0)}$ is initialized with degree, weighted degree, and the absolute weighted degree. The edge features $e_{uv}^{(0)}$ for  edge $(u, v)$ contain the edge weights and the absolute edge weights.

The GNN model is designed to learn the representations of both nodes and edges simultaneously across multiple layers via the message passing scheme. This model takes into account not only the features of nodes and edges but also their interactions, ensuring that both nodes and edges evolve over time as the network processes information. Let $\mathbf{H}^{(\ell)}$ denote the set of node representations $h_v^{(\ell)}$ and $\mathbf{E}^{(\ell)}$ denote the set of edge representations $e_{uv}^{(\ell)}$ at layer $\ell$. At layer $\ell$, the message passing scheme updates each node representation based on the neighboring nodes' representations at the previous layer $\ell - 1$; meanwhile, each edge representation is updated based on the edge's two corresponding nodes. Formally, we have:
\begin{align*}
h_v^{(\ell)} &=  \text{MLP}_1 \left(  h_v^{(\ell-1)} \oplus \sum_{u\in \mathcal{N}(v)}  h_{u}^{(\ell-1)} \oplus\sum_{u\in \mathcal{N}(v)} e_{v, u}^{(\ell-1)} \right),\\
e_{uv}^{(\ell)} &=  \text{MLP}_2\left(h_u^{(\ell-1)} \oplus h_v^{(\ell-1)} \oplus e_{uv}^{(\ell-1)}\right),
\end{align*}
where  $\text{MLP}(\cdot)$ denotes a Multilayer Perceptron\footnote{The two MLPs encoding for nodes and edges do not share parameters and are denoted as $\text{MLP}_1$ and $\text{MLP}_2$, respectively.}, $\mathcal{N}(v)$ denotes the set of neighboring nodes of $v$, and $\oplus$ denotes the operation of vector concatenation. For effective computation of combinatorial properties in the underlying graph, our MLP uses simple ReLU activation functions after each hidden linear transformation layer and a simple linear (identity) activation for the output layer.

An important step in our reduction framework is to predict for if each edge $(u, v)$ is aligned or anti-alignment, i.e., whether it should be merged or flip-merged, respectively. We leverage a logistic regression model for this task. Let $L$ denote the number of layers of message passing, we concatenate the node and edge representations of the last layer into a feature vector  $z_{uv}$
$$z_{uv} = h_u^{(L)} \oplus h_v^{(L)} \oplus e_{uv}^{(L)}.$$
The  logistic regression model predicts the probability on each edge (i.e. edge confidence) as:
\begin{equation}\label{eq:w_prefiction}
    \hat{y}_{uv} = \sigma(\langle w, z_{uv} \rangle),
\end{equation}
where $\sigma(\cdot)$ denotes the sigmoid function, $w$ is a learnable vector with equal length as the concatenated edge vector $z_{uv}$, and $\langle \cdot, \cdot \rangle$ denotes the inner product.

\subsection{Hybrid Loss Function for Iterative Contraction}

To simplify the Ising Hamiltonian, our method iteratively contracts one edge per iteration. The process involves combining a binary cross-entropy (BCE) loss for edge classification (alignment or anti-alignment) with confidence-based weighting to prioritize the most certain predictions.

Consider a graph with \( N \) edges, represented by the set \( \mathbf{L} = \{l_1, l_2, \dots, l_N\} \), where each edge \( l_i \) has a logit \( \hat{y}_{i} \) corresponding to the $i$th edge \( (u, v) \) and a ground-truth label \( y_{i} \in \{0, 1\} \) indicating alignment or anti-alignment, respectively. Neutral links are excluded from the loss computation since their processing (merge or flip-merge) preserves Ising optimality.

The confidence-based weight for each edge \( l_i \) is computed directly from \( \hat{y}_{i} \) as:
\[
c_i = \frac{\exp\left(|\hat{y}_{i} - 0.5| / T\right)}{\sum_{j=1}^{N} \exp\left(|\hat{y}_{j} - 0.5| / T\right)},
\]
where \( T \) is a temperature hyper-parameter. The absolute difference \(|\hat{y}_{i} - 0.5|\) measures the confidence of the prediction, with larger values indicating greater certainty. The softmax normalization ensures a smooth, differentiable prioritization of edges for contraction.

The weighting mechanism introduces a trade-off parameter, \(\lambda \in [0, 1]\), to combine the confidence-based weights \( c_i \) with uniform weights \footnote{An alternative formulation for \( c_i \) can be defined as \( c_i = \frac{(\hat{y}_{i} - 0.5)^p}{\sum_{j=1}^N ((\hat{y}_{j} - 0.5)^p} \), where even integer $p\geq 2$ is a parameter controlling the sensitivity of the confidence scores.}. The final weight for each edge is:
\[
w_i = \lambda \cdot c_i + (1 - \lambda).
\]

The hybrid loss function is defined as:
\[
\mathcal{L} = \sum_{i=1}^{N} w_i \cdot \left( -y_i \log(\hat{y}_i) - (1 - y_i) \log(1 - \hat{y}_i) \right).
\]

This interpolation enables different loss formulations:
\begin{itemize}
    \item \(\lambda = 0\): The loss reduces to standard binary cross-entropy (BCE), treating all edges equally.
    \item \(\lambda = 1\): The loss relies entirely on the confidence-based softmax weights, emphasizing high-confidence edges.
\end{itemize}

This formulation provides a flexible mechanism to adapt the loss to varying levels of confidence in predictions, balancing attention on certain and uncertain edges for effective edge contraction.

\section{Experiments} \label{sec:experiments}
We conducted extensive experiments to evaluate GRANITE's effectiveness in compressing Ising models while maintaining solution quality. Our evaluation focused on the trade-off between compression levels and solution accuracy across various graph topologies and sizes. The experiments were performed on D-Wave's Advantage Quantum Processing Unit (QPU), featuring the advanced Pegasus topology ($P_{16}$) with 5,640 qubits.
\subsection{Experimental Setup}

\begin{table*}[!h]
\small
\centering
\begin{tabular}{|c|c|r|r|r|r|r|}
\hline
\multirow{2}{*}{\textbf{Topology}} & \multirow{2}{*}{\textbf{Reduction (\%)}} & \multicolumn{5}{c|}{\textbf{n}} \\
\cline{3-7}
 &  & \multicolumn{1}{c|}{\textbf{25}} & \multicolumn{1}{c|}{\textbf{50}} & \multicolumn{1}{c|}{\textbf{100}} & \multicolumn{1}{c|}{\textbf{200}} & \multicolumn{1}{c|}{\textbf{400}} \\
\hline
\multirow{5}{*}{Erdős-Rényi} & \cellcolor{lightgray}\begin{tabular}[c]{@{}c@{}}0.0\%\\ Original Ising\end{tabular} & \cellcolor{lightgray}100.00 $\pm$ 0.00 & \cellcolor{lightgray}100.00 $\pm$ 0.00 & \cellcolor{lightgray}99.68 $\pm$ 0.19 & \cellcolor{lightgray}96.77 $\pm$ 0.45 & \cellcolor{lightgray} NaN\\
\cline{2-7}
 & 12.5\% & 98.21 $\pm$ 1.09 & 98.67 $\pm$ 0.44 & 96.70 $\pm$ 0.55 & 94.87 $\pm$ 0.37 & 89.36 $\pm$ 0.37 \\
 & 25.0\% & 96.03 $\pm$ 1.44 & 97.12 $\pm$ 0.76 & 94.26 $\pm$ 0.56 & 93.38 $\pm$ 0.54 & 88.58 $\pm$ 0.80 \\
 & 50.0\% & 94.58 $\pm$ 1.64 & 94.19 $\pm$ 1.42 & 91.23 $\pm$ 1.00 & 91.56 $\pm$ 0.63 & 88.02 $\pm$ 0.63 \\
 & 75.0\% & 93.20 $\pm$ 1.55 & 92.26 $\pm$ 1.40 & 89.60 $\pm$ 0.97 & 91.07 $\pm$ 0.82 & 88.17 $\pm$ 0.79 \\
\hline
\multirow{5}{*}{Barabási-Albert} & \cellcolor{lightgray}\begin{tabular}[c]{@{}c@{}}0.0\%\\ Original Ising\end{tabular} & \cellcolor{lightgray}100.00 $\pm$ 0.00 & \cellcolor{lightgray}100.00 $\pm$ 0.00 & \cellcolor{lightgray}99.93 $\pm$ 0.07 & \cellcolor{lightgray}97.25 $\pm$ 0.25 & \cellcolor{lightgray}89.11 $\pm$ 0.48 \\
\cline{2-7}
 & 12.5\% & 97.69 $\pm$ 1.22 & 97.54 $\pm$ 0.91 & 97.79 $\pm$ 0.65 & 94.71 $\pm$ 0.80 & 88.53 $\pm$ 1.11 \\
 & 25.0\% & 97.12 $\pm$ 1.38 & 95.63 $\pm$ 1.24 & 97.09 $\pm$ 0.75 & 92.69 $\pm$ 1.08 & 87.84 $\pm$ 1.11 \\
 & 50.0\% & 94.88 $\pm$ 1.59 & 93.00 $\pm$ 1.45 & 94.87 $\pm$ 1.14 & 92.66 $\pm$ 0.92 & 88.76 $\pm$ 0.60 \\
 & 75.0\% & 92.43 $\pm$ 1.88 & 91.89 $\pm$ 1.21 & 93.37 $\pm$ 1.28 & 92.30 $\pm$ 0.87 & 88.72 $\pm$ 0.96 \\
\hline
\multirow{5}{*}{Watts-Strogatz} & \cellcolor{lightgray}\begin{tabular}[c]{@{}c@{}}0.0\%\\ Original Ising\end{tabular} & \cellcolor{lightgray}100.00 $\pm$ 0.00 & \cellcolor{lightgray}100.00 $\pm$ 0.00 & \cellcolor{lightgray}100.00 $\pm$ 0.00 & \cellcolor{lightgray}99.14 $\pm$ 0.14 & \cellcolor{lightgray}96.57 $\pm$ 0.26 \\
\cline{2-7}
 & 12.5\% & 98.06 $\pm$ 1.13 & 98.85 $\pm$ 0.52 & 99.08 $\pm$ 0.40 & 96.73 $\pm$ 0.59 & 91.96 $\pm$ 0.72 \\
 & 25.0\% & 95.95 $\pm$ 1.30 & 96.86 $\pm$ 0.90 & 97.40 $\pm$ 0.52 & 96.28 $\pm$ 0.56 & 90.73 $\pm$ 1.28 \\
 & 50.0\% & 93.39 $\pm$ 1.08 & 96.26 $\pm$ 1.07 & 95.13 $\pm$ 0.83 & 94.69 $\pm$ 0.41 & 90.63 $\pm$ 0.94 \\
 & 75.0\% & 91.69 $\pm$ 1.31 & 94.69 $\pm$ 1.22 & 92.74 $\pm$ 1.26 & 92.23 $\pm$ 0.53 & 91.03 $\pm$ 0.88 \\
\hline
\end{tabular}
\caption{Solution optimality on D-Wave quantum annealers after and before compressing Ising models with GRANITE.}
\label{tab:granite_optimality}
\end{table*}

\begin{figure}[H]
    \centering
    \includegraphics[width=0.45\textwidth]{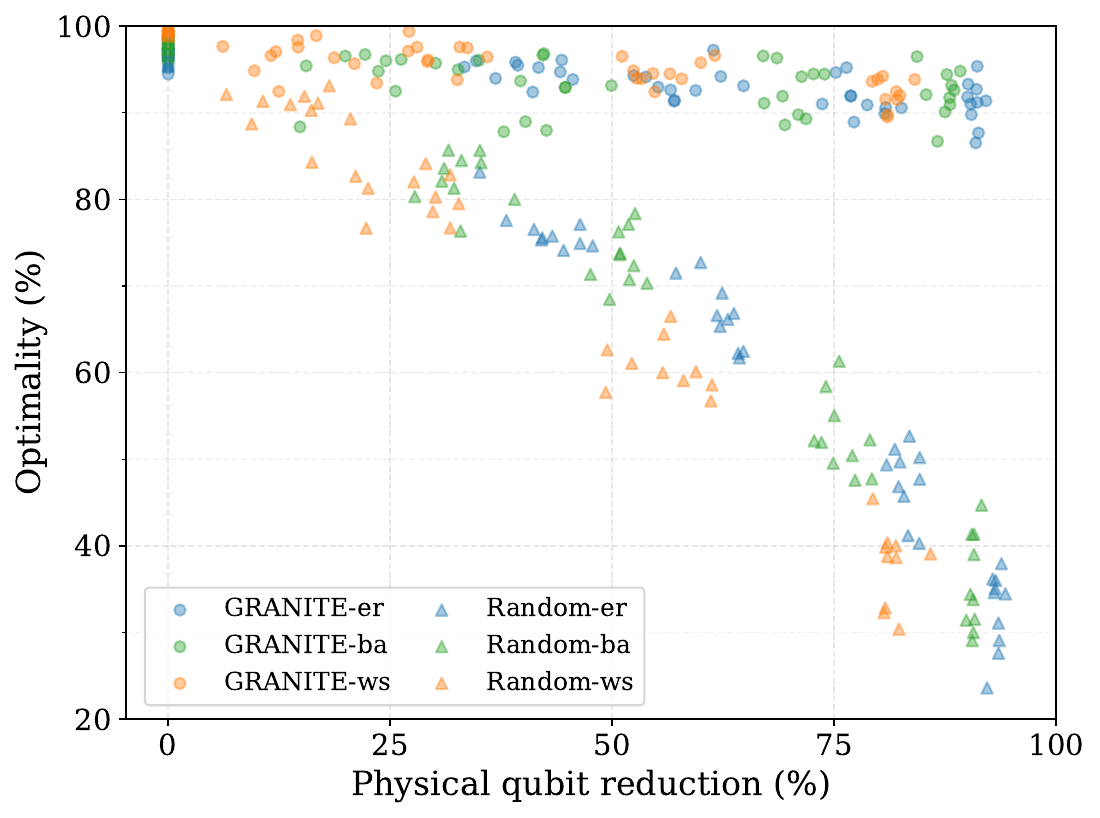}
    \caption{GRANITE vs. random, the random merge and flip-merge of edge for n = 200 across three different topologies.}
    \label{fig:granite-random}
\end{figure}

\begin{table*}[!h]
\renewcommand{\arraystretch}{1.2}
\centering
\small
\begin{tabular}{|c|c|r|r|r|r|r|}
\hline
\multirow{2}{*}{\textbf{Topology}} & \multirow{2}{*}{\textbf{Edge reduction (\%)}} & \multicolumn{5}{c|}{\textbf{n}} \\
\cline{3-7}
 &  & \textbf{25} & \textbf{50} & \textbf{100} & \textbf{200} & \textbf{400} \\
\hline
\multirow{6}{*}{Erdős-Rényi} & \cellcolor[rgb]{0.902,0.902,0.902}0.0\% & \cellcolor[rgb]{0.902,0.902,0.902}47.6 qubits & \cellcolor[rgb]{0.902,0.902,0.902}133.2 qubits& \cellcolor[rgb]{0.902,0.902,0.902}414.3 qubits& \cellcolor[rgb]{0.902,0.902,0.902}1362.1 qubits& \cellcolor[rgb]{0.902,0.902,0.902} $>$ 5760 qubits \\
 & \cellcolor[rgb]{0.902,0.902,0.902}Original Ising & \cellcolor[rgb]{0.902,0.902,0.902}100\% & \cellcolor[rgb]{0.902,0.902,0.902}100\% & \cellcolor[rgb]{0.902,0.902,0.902}100\% & \cellcolor[rgb]{0.902,0.902,0.902}100\% & \cellcolor[rgb]{0.902,0.902,0.902}100\% \\
\cline{2-7}
 & 12.5\% & 82.1\% & 71.8\% & 67.3\% & 59.8\% & $<$ 46.7\% \\
 & 25.0\% & 68.3\% & 56.3\% & 49.3\% & 42.0\% & $<$ 32.2\% \\
 & 50.0\% & 45.0\% & 33.1\% & 27.4\% & 22.1\% & $<$ 16.3\% \\
 & 75.0\% & 24.2\% & 17.0\% & 12.3\% & 9.1\% & $<$ 5.3\% \\
\hline
\multirow{6}{*}{Barabási-Albert} & \cellcolor[rgb]{0.902,0.902,0.902}0.0\% & \cellcolor[rgb]{0.902,0.902,0.902}39.7 qubits& \cellcolor[rgb]{0.902,0.902,0.902}110.0 qubits& \cellcolor[rgb]{0.902,0.902,0.902}327.6 qubits& \cellcolor[rgb]{0.902,0.902,0.902}1008.8 qubits& \cellcolor[rgb]{0.902,0.902,0.902}3621.2 qubits\\
 & \cellcolor[rgb]{0.902,0.902,0.902}Original Ising & \cellcolor[rgb]{0.902,0.902,0.902}100\% & \cellcolor[rgb]{0.902,0.902,0.902}100\% & \cellcolor[rgb]{0.902,0.902,0.902}100\% & \cellcolor[rgb]{0.902,0.902,0.902}100\% & \cellcolor[rgb]{0.902,0.902,0.902}100\% \\
\cline{2-7}
 & 12.5\% & 82.9\% & 79.7\% & 79.5\% & 76.3\% & 68.5\% \\
 & 25.0\% & 74.1\% & 63.0\% & 60.4\% & 58.0\% & 48.9\% \\
 & 50.0\% & 46.3\% & 39.0\% & 35.1\% & 29.8\% & 22.2\% \\
 & 75.0\% & 28.5\% & 20.3\% & 16.2\% & 12.7\% & 7.4\% \\
\hline
\multirow{6}{*}{Watts-Strogatz} & \cellcolor[rgb]{0.902,0.902,0.902}0.0\% & \cellcolor[rgb]{0.902,0.902,0.902}43.2 qubits & \cellcolor[rgb]{0.902,0.902,0.902}89.9 qubits & \cellcolor[rgb]{0.902,0.902,0.902}223.7 qubits& \cellcolor[rgb]{0.902,0.902,0.902}578.2 qubits& \cellcolor[rgb]{0.902,0.902,0.902}1544.5 qubits\\
 & \cellcolor[rgb]{0.902,0.902,0.902}Original Ising & \cellcolor[rgb]{0.902,0.902,0.902}100\% & \cellcolor[rgb]{0.902,0.902,0.902}100\% & \cellcolor[rgb]{0.902,0.902,0.902}100\% & \cellcolor[rgb]{0.902,0.902,0.902}100\% & \cellcolor[rgb]{0.902,0.902,0.902}100\% \\
\cline{2-7}
 & 12.5\% & 84.7\% & 88.7\% & 84.8\% & 85.3\% & 87.2\% \\
 & 25.0\% & 72.0\% & 74.0\% & 72.5\% & 70.6\% & 70.7\% \\
 & 50.0\% & 46.5\% & 48.7\% & 47.0\% & 44.4\% & 43.0\% \\
 & 75.0\% & 25.2\% & 24.5\% & 20.1\% & 18.6\% & 16.5\% \\
\hline
\end{tabular}
\caption{Qubit reduction on D-Wave quantum annealers before and after compressing Ising models with GRANITE}
\label{tab:granite_reduction}
\end{table*}

\begin{table}[t]
\centering
\small
\setlength{\tabcolsep}{4pt}
\begin{tabular}{|l|c|c|c|c|}
\hline
& BCE & MSE & Hybrid & Softmax \\
\hline
ER & 86.90 $\pm$ 1.08 & 75.68 $\pm$ 3.85 & \textbf{91.07 $\pm$ 0.82} & 89.68 $\pm$ 1.07 \\
BA & 89.05 $\pm$ 1.35 & 92.07 $\pm$ 0.69 & \textbf{92.30 $\pm$ 0.87} & 78.89 $\pm$ 2.44 \\
WS & 92.58 $\pm$ 0.30 & \textbf{94.02 $\pm$ 0.60} & 92.23 $\pm$ 0.53 & 92.60 $\pm$ 0.51 \\
\hline
\end{tabular}
\caption{Optimality across different loss functions for n=200, edge reduction 75\%.}
\label{tab:comparison-optimality-loss-function}
\end{table}

\begin{table}[t]
\centering
\small
\setlength{\tabcolsep}{4pt}
\begin{tabular}{|l|c|c|c|}
\hline
\# Layers & ER & BA & WS \\
\hline
1 & 76.64 $\pm$ 0.86 & 82.95 $\pm$ 0.72 & 85.73 $\pm$ 1.05 \\
2 & 88.09 $\pm$ 0.45 & 86.47 $\pm$ 1.07 & 91.56 $\pm$ 1.03 \\
3 & \textbf{91.07 $\pm$ 0.82} & \textbf{92.30 $\pm$ 0.87} & \textbf{92.23 $\pm$ 0.53} \\
4 & 81.07 $\pm$ 1.27 & 86.47 $\pm$ 0.91 & 85.79 $\pm$ 1.39 \\
5 & 84.48 $\pm$ 1.49 & 90.72 $\pm$ 0.85 & 90.94 $\pm$ 0.63 \\
\hline
\end{tabular}
\caption{Optimality with different number of GNN layers at n = 200 and reduction rate = 75\%.}
\label{tab:comparison-optimality-layers}
\end{table}

\paragraph{Dataset.} We generate random Ising Hamiltonians represented as graphs with the spins as nodes and edges following three  graph topologies Erdős-Rényi (ER), Barabási-Albert (BA), and Watts-Strogatz (WS) models. Nodes have zero linear biases, i.e., $h_i = 0$ for all $i$ and the edge weight $J_{ij}$  sampled uniformly randomly in the range $(-5,5)$.

The dataset includes 97,500 graphs (325 distinct configurations × 100 instances × 3 topologies), split into 80\% training and 20\% validation sets.For each type of topology, the number of nodes ranges from 2 to 26 (25 distinct sizes), while the average node degree varies from 1 to $n - 1$ (sum up to 325 instances). For each combination of node count and average degree, we generate 100 unique graphs, resulting in a total dataset of 97,500 graphs (325 $\times$ 100 instances $\times$ 3 types of topologies). We split training at a ratio of 80\%/20\% for training/validation respectively. The evaluation set contains 20,000 graphs to assess the effectiveness of unseen graphs. The ground state labels for each edge were computed using exhaustive search, indicating whether the edge could be \textit{Alignment} (category A), \textit{Anti-alignment}  (category C), or required further processing \textit{Neutral} (category B).

\subsubsection{Hyperparameters.}
We employ the Adam optimizer with a learning rate of 0.001. The maximum number of iterations is set to 300, with the best model saved based on the lowest loss performance on the validation set. The size of the $\text{MLP}$ layers is relatively small, depending on the size of $\mathbf{H}^{(0)}$ and $\mathbf{E}^{(0)}$, which in this case yields 2 and 2, respectively. By default, the GRANITE consists of  three GNN layers and use a hybrid loss function with $\lambda = 0.5$.

\subsubsection{Environment.}
We conducted experiments on the D-Wave Advantage 4.1 Quantum Processing Unit (QPU), utilizing the advanced Pegasus topology with 5,640 qubits and an annealing time of 40 ns. The minor-embedding of the Ising Hamiltonians are found with minorminer \cite{cai2014practical}. Optimal solutions for each Ising instance is found using the Gurobi Optimizer \cite{gurobi}.

\subsubsection{Evaluation metrics}
We report two metrics that measure the solution quality and the qubit reduction levels. The solution quality is measured using optimality, the ratio between the best energy found using D-wave quantum annealer, denoted by $E_{\text{best}}$, and the minimum energy found using Gurobi, denoted by $E_{\text{min}}$. Formally, the optimality is computed as
\begin{equation}
\text{Optimality(\%)} = 1 - \frac{|E_{\text{best}}- E_{\text{min}}|}{|E_{\text{min}}|},
\end{equation}
to accommodate for the case when $E_{\text{best}} > 0$ (and $E_{\text{min}} < 0$). When optimal solutions are found, the optimality will be one.

The qubit reduction is measured as 
\[
\text{reduction} = 1 - \frac{q_{\text{compressed}}}{q_{\text{original}}},
\]
where $q_{\text{compressed}}$ represents the number of physical qubits after compression and $q_{\text{original}}$ represents the original number of physical qubits before compression. 

\subsection{Experiment results}

\paragraph{Solution quality.} Our experimental results demonstrate GRANITE's effectiveness in compressing Ising models while maintaining high solution quality across different graph types and sizes. Tables \ref{tab:granite_optimality} and \ref{tab:granite_reduction} present optimality and qubit reductions for different compression ratios (12.5\%, 25\%, 50\%, 75\%) across graph of sizes $n=\{25, 50, 100, 200, 400\}$.
The solution quality remains high across all graph types, even at aggressive compression levels. For instance, with 75\% edge reduction on graphs with $n = 200$, we maintain optimality above 91\% for all three topologies. We observe a slight degradation in solution quality as graph size increases.

\paragraph{Qubit Reduction.} For the largest Ising instances ($n = 400$), the Ising Hamiltonian exceeds the hardware constraint and cannot be solved on the D-Wave quantum annealer for Erdős-Rényi  model while taking 3621 and 1545 qubits on average for Barabási-Albert, and Watts-Strogatz instances. At 75\% edge reduction, the remaining physical qubit ratios reach as low as 5.3\% (Erdős-Rényi), 7.4\% (Barabási-Albert), and 16.5\% (Watts-Strogatz) of the original number of qubits.

\paragraph{Comparison with Baselines.} There is a clear performance gap between GRANITE and random reduction strategies as shown in Figure~\ref{fig:granite-random}. While GRANITE maintains optimality above 91\% across all topologies, random baseline approaches achieve significantly lower performance, under 40\% for all network topologies. This stark contrast demonstrates that GRANITE's learned compression strategies significantly outperform random reduction approaches. Moreover, as the first method to offer variable compression ratios, GRANITE provides a unique advantage over existing approaches, which typically achieve less than 20\% reduction on average.

\paragraph{Ablation Studies.} We conducted comprehensive ablation studies to evaluate the impact of two key architectural choices: loss function and the number of GNN layers. For loss functions, we compared Binary Cross-Entropy (BCE), Mean Squared Error (MSE), our proposed hybrid loss, and softmax-based weighting. The results in Table \ref{tab:comparison-optimality-loss-function} demonstrate that our hybrid loss function achieves better performance on both Erdős-Rényi (91.07\%) and Barabási-Albert (92.30\%) graphs, while performing competitively on Watts-Strogatz topology (92.23\%).  GNN depth analysis, shown in Table \ref{tab:comparison-optimality-layers}, reveals that a three-layer architecture consistently achieves optimal performance across all graph types, with peak optimality of 91.07\% (ER), 92.30\% (BA), and 92.23\% (WS). Deeper GNN architecture with 4 and 5 layers shows performance degradation, suggesting that three layers provide sufficient representational capacity while avoiding overfitting. These findings guided our choice of hybrid loss and three-layer architecture for the final GRANITE model.

These results demonstrate that our GRANITE model effectively compresses Ising Hamiltonians while maintaining high solution quality. The model achieves substantial qubit reduction, particularly for larger graphs, which is crucial for making quantum annealing more accessible for solving complex optimization problems.

The consistent performance across different graph types suggests that our approach is robust and generalizable to various network structures. This is particularly important as real-world optimization problems often involve diverse graph topologies.

\section{Conclusion and Future Work} \label{sec:conclusion}
While extensive efforts have been made to tackle combinatorial optimization problems from various angles, particularly in solver development, the approach of dynamic qubit compression has remained largely unexplored. This paper introduces GRANITE, an automated, fast system that iteratively compresses large Ising models while maintaining solution accuracy.
GRANITE's has demonstrated effective compression of large-scale graphs across different random graph models, resulting in significant physical qubit reduction on D-Wave quantum annealing machines.
The potential of GRANITE opens up new avenues for solving large-scale optimization problems on quantum hardware with limited qubit counts. Future work will focus on investigating the effectiveness of our qubit compression GRANITE model on real-world large graphs and exploring potential applications in various domains.

Future work will focus on investigating the effectiveness of our qubit compression model on real-world large graphs and exploring potential applications in various domains. Additionally, adapting GRANITE to other quantum computing technologies while addressing hardware-specific noise challenges presents an important direction for extending its impact across different quantum computing paradigms.

\newpage
\section{Acknowledgment}
This work was supported in part by NSF  AMPS-2229075, VCU Quest Award, and  Commonwealth Cyber Initiative (CCI) awards.

\newpage
\appendix
\appendix
\section{Proofs of Theorem~\ref{theo:co-np-hard}}
We study the computational complexity of determining spin equivalence in ground states of Ising Hamiltonians. Our main result shows that this problem is Co-NP-hard through a reduction from a satisfiability variant that we prove to be Co-NP-complete.

\begin{definition}[TAUTOLOGY]
The TAUTOLOGY problem takes as input a Boolean formula $\psi$ and determines whether $\psi$ evaluates to true under all possible truth assignments. This problem is Co-NP-complete as its complement, SAT, is NP-complete.
\end{definition}

\begin{definition}[All-SAT-EQUAL]
The All-SAT-EQUAL problem takes as input a Boolean formula $\phi$ in Conjunctive Normal Form (CNF) and two variables $x_i$ and $x_j$, and determines whether all satisfying assignments of $\phi$ satisfy $x_i = x_j$.
\end{definition}

\begin{lemma}
All-SAT-EQUAL is Co-NP-complete.
\end{lemma}

\begin{proof}
For membership in Co-NP, observe that its complement (existence of a satisfying assignment where $x_i \neq x_j$) is in NP, as such an assignment can be verified in polynomial time.

For hardness, we reduce from TAUTOLOGY. Given formula $\psi$, construct $\phi = \psi \land (x_i \leftrightarrow x_j)$. In CNF, this biconditional is represented as \((x_i \lor \neg x_j) \land (\neg x_i \lor x_j)\). Then $\psi$ is a tautology if and only if all satisfying assignments of $\phi$ satisfy $x_i = x_j$. The reduction is polynomial-time, establishing Co-NP-hardness.
\end{proof}

To establish our main result, we first prove a folklore result that provides a binary quadratic encoding of 3-SAT clauses.

\begin{lemma}[3-SAT Clause Encoding]
For a SAT clause $C_i$ with literals $x_1, x_2, x_3 \in \{0, 1\}$, define:
\begin{align*}
Q(C_i) = & \; x_c\big(2 - (x_1 + x_2 + x_3)\big) \\
         & \; + (x_1x_2 + x_2x_3 + x_3x_1)
          - (x_1 + x_2 + x_3) + 1,
\end{align*}
where $x_c \in \{0, 1\}$ is an auxiliary variable. Then $\min Q(C_i) = 0$ if and only if $C_i$ is satisfied, and $\min Q(C_i) = 1$ otherwise.
\end{lemma}

\begin{proof}
Let $s = x_1 + x_2 + x_3$. For Boolean variables, the term $x_1x_2 + x_2x_3 + x_3x_1$ equals $s$ when $s = 3$, equals $0$ when $s = 1$, and equals $1$ when $s = 2$. Thus:
\[ Q(C_i) = \begin{cases}
2x_c + 1 & \text{if } s = 0 \\
x_c(2-s) & \text{if } s = 1 \text{ or } 2 \\
x_c(-1) + 1 & \text{if } s = 3
\end{cases} \]

When $s = 0$ (unsatisfied clause), $\min Q(C_i) = 1$ with $x_c = 0$. When $s = 1 \text{ or } 2$ (satisfied clause), $x_c(2-s) \geq 0$, so $\min Q(C_i) = 0$ with $x_c = 0$. When $s = 3$ (satisfied clause), $\min Q(C_i) = 0$ with $x_c = 1$.
\end{proof}

We now state and prove our main result on the Co-NP-hardness of
determining whether two spins $s_i = s_j$ are equal in all ground states of an Ising Hamiltonian.

\begin{proof}
We reduce from All-SAT-EQUAL. Given an instance $(\phi, x_i, x_j)$, construct a global QUBO function:
\[ Q(\phi) = \sum_k Q(C_k) \]
where the sum is over all clauses $C_k$ in $\phi$. By the previous lemma, $\min Q(\phi) = 0$ if and only if $\phi$ is satisfied.

Transform $Q(\phi)$ to an Ising Hamiltonian $H$ by substituting $x_k = \frac{1 + s_k}{2}$ for each variable $x_k$. This mapping preserves the equivalence between satisfying assignments and ground states. The linear relationship between variables ensures $x_i = x_j$ if and only if $s_i = s_j$.

Each clause requires one auxiliary variable $x_c$, and the transformation preserves polynomial size with integer coefficients. Therefore, determining whether $s_i = s_j$ in all ground states of $H$ is equivalent to determining whether $x_i = x_j$ in all satisfying assignments of $\phi$, establishing Co-NP-hardness.
\end{proof}

This result has implications for the computational complexity of analyzing ground state properties in Ising systems, suggesting that even seemingly simple questions about spin equivalence can be computationally intractable.


\begin{thebibliography}{59}
\providecommand{\natexlab}[1]{#1}

\bibitem[{Angelini and Ricci-Tersenghi(2022)}]{Angelini2022}
Angelini, M.~C.; and Ricci-Tersenghi, F. 2022.
\newblock Modern graph neural networks do worse than classical greedy
  algorithms in solving combinatorial optimization problems like maximum
  independent set.
\newblock \emph{Nature Machine Intelligence}, 29--31.

\bibitem[{Angelini and Ricci-Tersenghi(2023)}]{angelini2023modern}
Angelini, M.~C.; and Ricci-Tersenghi, F. 2023.
\newblock Modern graph neural networks do worse than classical greedy
  algorithms in solving combinatorial optimization problems like maximum
  independent set.
\newblock \emph{Nature Machine Intelligence}, 5(1): 29--31.

\bibitem[{Arute et~al.(2019)Arute, Arya, Babbush, Bacon, Bardin, Barends,
  Biswas, Boixo, Brandao, Buell et~al.}]{arute2019quantum}
Arute, F.; Arya, K.; Babbush, R.; Bacon, D.; Bardin, J.~C.; Barends, R.;
  Biswas, R.; Boixo, S.; Brandao, F.~G.; Buell, D.~A.; et~al. 2019.
\newblock Quantum supremacy using a programmable superconducting processor.
\newblock \emph{Nature}, 574(7779): 505--510.

\bibitem[{Bauza and Lidar(2024)}]{bauza2024scaling}
Bauza, H.~M.; and Lidar, D.~A. 2024.
\newblock Scaling Advantage in Approximate Optimization with Quantum Annealing.
\newblock arXiv:2401.07184.

\bibitem[{Boettcher(2022)}]{Boettcher2022}
Boettcher, S. 2022.
\newblock Inability of a graph neural network heuristic to outperform greedy
  algorithms in solving combinatorial optimization problems.
\newblock \emph{Nature Machine Intelligence}, 2522--5839.

\bibitem[{Boothby et~al.(2020)Boothby, Bunyk, Raymond, and Roy}]{Boothby2020}
Boothby, K.; Bunyk, P.; Raymond, J.; and Roy, A. 2020.
\newblock {Next-Generation Topology of D-Wave Quantum Processors}.
\newblock ArXiv:2003.00133.

\bibitem[{Boros and Hammer(2002)}]{boros2002pseudo}
Boros, E.; and Hammer, P.~L. 2002.
\newblock Pseudo-boolean optimization.
\newblock \emph{Discrete applied mathematics}, 123(1-3): 155--225.

\bibitem[{Cai, Macready, and Roy(2014)}]{cai2014practical}
Cai, J.; Macready, W.~G.; and Roy, A. 2014.
\newblock A practical heuristic for finding graph minors.
\newblock \emph{arXiv preprint arXiv:1406.2741}.

\bibitem[{Carleo et~al.(2019)Carleo, Cirac, Cranmer, Daudet, Schuld, Tishby,
  Vogt-Maranto, and Zdeborov\'{a}}]{Carleo2019}
Carleo, G.; Cirac, I.; Cranmer, K.; Daudet, L.; Schuld, M.; Tishby, N.;
  Vogt-Maranto, L.; and Zdeborov\'{a}, L. 2019.
\newblock Machine learning and the physical sciences.
\newblock \emph{Rev. Mod. Phys.}, 91: 045002.

\bibitem[{Choi(2008)}]{choi2008minor}
Choi, V. 2008.
\newblock Minor-embedding in adiabatic quantum computation: I. The parameter
  setting problem.
\newblock \emph{Quantum Information Processing}, 7(5): 193--209.

\bibitem[{Choi(2011)}]{choi2011minor}
Choi, V. 2011.
\newblock Minor-embedding in adiabatic quantum computation: II. Minor-universal
  graph design.
\newblock \emph{Quantum Information Processing}, 10(3): 343--353.

\bibitem[{Dai et~al.(2017)Dai, Khalil, Zhang, Dilkina, and Song}]{Dai2017LC}
Dai, H.; Khalil, E.~B.; Zhang, Y.; Dilkina, B.; and Song, L. 2017.
\newblock Learning Combinatorial Optimization Algorithms over Graphs.
\newblock In \emph{Proceedings of the 31st International Conference on Neural
  Information Processing Systems}, NIPS'17, 6351–6361.
\newblock ISBN 9781510860964.

\bibitem[{Farhi et~al.(2000)Farhi, Goldstone, Gutmann, and Sipser}]{Farhi2000}
Farhi, E.; Goldstone, J.; Gutmann, S.; and Sipser, M. 2000.
\newblock {Quantum Computation by Adiabatic Evolution}.
\newblock ArXiv:quant-ph/0001106.

\bibitem[{Fox, Branson, and Walker(2021)}]{fox2021mrna}
Fox, D.~M.; Branson, K.~M.; and Walker, R.~C. 2021.
\newblock mRNA codon optimization with quantum computers.
\newblock \emph{PloS one}, 16(10): e0259101.

\bibitem[{Fox et~al.(2022)Fox, MacDermaid, Schreij, Zwierzyna, and
  Walker}]{fox2022rna}
Fox, D.~M.; MacDermaid, C.~M.; Schreij, A.~M.; Zwierzyna, M.; and Walker, R.~C.
  2022.
\newblock RNA folding using quantum computers.
\newblock \emph{PLOS Computational Biology}, 18(4): e1010032.

\bibitem[{Gary and Johnson(1979)}]{gary1979computers}
Gary, M.~R.; and Johnson, D.~S. 1979.
\newblock Computers and Intractability: A Guide to the Theory of
  NP-completeness.

\bibitem[{Gasse et~al.(2019)Gasse, Chetelat, Ferroni, Charlin, and
  Lodi}]{Gasse2019}
Gasse, M.; Chetelat, D.; Ferroni, N.; Charlin, L.; and Lodi, A. 2019.
\newblock Exact Combinatorial Optimization with Graph Convolutional Neural
  Networks.
\newblock In \emph{Neural Information Processing Systems}.

\bibitem[{Glover, Kochenberger, and Du(2018)}]{glover2018tutorial}
Glover, F.; Kochenberger, G.; and Du, Y. 2018.
\newblock A tutorial on formulating and using QUBO models.
\newblock \emph{arXiv preprint arXiv:1811.11538}.

\bibitem[{Goto et~al.(2021)Goto, Endo, Suzuki, Sakai, Kanao, Hamakawa, Hidaka,
  Yamasaki, and Tatsumura}]{goto2021high}
Goto, H.; Endo, K.; Suzuki, M.; Sakai, Y.; Kanao, T.; Hamakawa, Y.; Hidaka, R.;
  Yamasaki, M.; and Tatsumura, K. 2021.
\newblock High-performance combinatorial optimization based on classical
  mechanics.
\newblock \emph{Science Advances}, 7(6): eabe7953.

\bibitem[{Grozea et~al.(2021)Grozea, Hans, Koch, Riehn, and
  Wolf}]{grozea2021optimising}
Grozea, C.; Hans, R.; Koch, M.; Riehn, C.; and Wolf, A. 2021.
\newblock Optimising Rolling Stock Planning including Maintenance with
  Constraint Programming and Quantum Annealing.
\newblock \emph{arXiv preprint arXiv:2109.07212}.

\bibitem[{{Gurobi Optimization, LLC}(2024)}]{gurobi}
{Gurobi Optimization, LLC}. 2024.
\newblock {Gurobi Optimizer Reference Manual}.

\bibitem[{Hammer, Hansen, and Simeone(1984)}]{hammer1984roof}
Hammer, P.~L.; Hansen, P.; and Simeone, B. 1984.
\newblock Roof duality, complementation and persistency in quadratic 0--1
  optimization.
\newblock \emph{Mathematical programming}, 28(2): 121--155.

\bibitem[{Hamze et~al.(2020)Hamze, Raymond, Pattison, Biswas, and
  Katzgraber}]{hamze2020wishart}
Hamze, F.; Raymond, J.; Pattison, C.~A.; Biswas, K.; and Katzgraber, H.~G.
  2020.
\newblock Wishart planted ensemble: A tunably rugged pairwise Ising model with
  a first-order phase transition.
\newblock \emph{Physical Review E}, 101(5): 052102.

\bibitem[{Harris et~al.(2018)Harris, Sato, Berkley, Reis, Altomare, Amin,
  Boothby, Bunyk, Deng, Enderud et~al.}]{harris2018phase}
Harris, R.; Sato, Y.; Berkley, A.; Reis, M.; Altomare, F.; Amin, M.; Boothby,
  K.; Bunyk, P.; Deng, C.; Enderud, C.; et~al. 2018.
\newblock Phase transitions in a programmable quantum spin glass simulator.
\newblock \emph{Science}, 361(6398): 162--165.

\bibitem[{Hauke et~al.(2020)Hauke, Katzgraber, Lechner, Nishimori, and
  Oliver}]{Hauke_2020}
Hauke, P.; Katzgraber, H.~G.; Lechner, W.; Nishimori, H.; and Oliver, W.~D.
  2020.
\newblock Perspectives of quantum annealing: methods and implementations.
\newblock \emph{Reports on Progress in Physics}, 83(5): 054401.

\bibitem[{Hibat-Allah et~al.(2020)Hibat-Allah, Ganahl, Hayward, Melko, and
  Carrasquilla}]{hibat2020recurrent}
Hibat-Allah, M.; Ganahl, M.; Hayward, L.~E.; Melko, R.~G.; and Carrasquilla, J.
  2020.
\newblock Recurrent neural network wave functions.
\newblock \emph{Physical Review Research}, 2(2): 023358.

\bibitem[{Hibat-Allah et~al.(2021)Hibat-Allah, Inack, Wiersema, Melko, and
  Carrasquilla}]{hibat2021variational}
Hibat-Allah, M.; Inack, E.~M.; Wiersema, R.; Melko, R.~G.; and Carrasquilla, J.
  2021.
\newblock Variational neural annealing.
\newblock \emph{Nature Machine Intelligence}, 3(11): 952--961.

\bibitem[{Honjo et~al.(2021)Honjo, Sonobe, Inaba, Inagaki, Ikuta, Yamada,
  Kazama, Enbutsu, Umeki, Kasahara et~al.}]{honjo2021100}
Honjo, T.; Sonobe, T.; Inaba, K.; Inagaki, T.; Ikuta, T.; Yamada, Y.; Kazama,
  T.; Enbutsu, K.; Umeki, T.; Kasahara, R.; et~al. 2021.
\newblock 100,000-spin coherent Ising machine.
\newblock \emph{Science advances}, 7(40): eabh0952.

\bibitem[{Joshi, Laurent, and Bresson(2019)}]{Joshi2019AnEG}
Joshi, C.~K.; Laurent, T.; and Bresson, X. 2019.
\newblock An Efficient Graph Convolutional Network Technique for the Travelling
  Salesman Problem.
\newblock \emph{ArXiv}, abs/1906.01227.

\bibitem[{Kadowaki and Nishimori(1998{\natexlab{a}})}]{kadowaki1998quantum}
Kadowaki, T.; and Nishimori, H. 1998{\natexlab{a}}.
\newblock Quantum annealing in the transverse Ising model.
\newblock \emph{Physical Review E}, 58(5): 5355.

\bibitem[{Kadowaki and Nishimori(1998{\natexlab{b}})}]{Kadowaki_1998}
Kadowaki, T.; and Nishimori, H. 1998{\natexlab{b}}.
\newblock {Quantum annealing in the transverse Ising model}.
\newblock \emph{Physical Review E}, 58(5): 5355--5363.

\bibitem[{King et~al.(2018)King, Bernoudy, King, Berkley, and
  Lanting}]{king2018emulating}
King, A.~D.; Bernoudy, W.; King, J.; Berkley, A.~J.; and Lanting, T. 2018.
\newblock Emulating the coherent Ising machine with a mean-field algorithm.
\newblock \emph{arXiv preprint arXiv:1806.08422}.

\bibitem[{King et~al.(2024)King, Nocera, Rams, Dziarmaga, Wiersema, Bernoudy,
  Raymond, Kaushal, Heinsdorf, Harris, Boothby, Altomare, Berkley, Boschnak,
  Chern, Christiani, Cibere, Connor, Dehn, Deshpande, Ejtemaee, Farré, Hamer,
  Hoskinson, Huang, Johnson, Kortas, Ladizinsky, Lai, Lanting, Li, MacDonald,
  Marsden, McGeoch, Molavi, Neufeld, Norouzpour, Oh, Pasvolsky, Poitras,
  Poulin-Lamarre, Prescott, Reis, Rich, Samani, Sheldan, Smirnov, Sterpka,
  Clavera, Tsai, Volkmann, Whiticar, Whittaker, Wilkinson, Yao, Yi, Sandvik,
  Alvarez, Melko, Carrasquilla, Franz, and Amin}]{king2024computational}
King, A.~D.; Nocera, A.; Rams, M.~M.; Dziarmaga, J.; Wiersema, R.; Bernoudy,
  W.; Raymond, J.; Kaushal, N.; Heinsdorf, N.; Harris, R.; Boothby, K.;
  Altomare, F.; Berkley, A.~J.; Boschnak, M.; Chern, K.; Christiani, H.;
  Cibere, S.; Connor, J.; Dehn, M.~H.; Deshpande, R.; Ejtemaee, S.; Farré, P.;
  Hamer, K.; Hoskinson, E.; Huang, S.; Johnson, M.~W.; Kortas, S.; Ladizinsky,
  E.; Lai, T.; Lanting, T.; Li, R.; MacDonald, A. J.~R.; Marsden, G.; McGeoch,
  C.~C.; Molavi, R.; Neufeld, R.; Norouzpour, M.; Oh, T.; Pasvolsky, J.;
  Poitras, P.; Poulin-Lamarre, G.; Prescott, T.; Reis, M.; Rich, C.; Samani,
  M.; Sheldan, B.; Smirnov, A.; Sterpka, E.; Clavera, B.~T.; Tsai, N.;
  Volkmann, M.; Whiticar, A.; Whittaker, J.~D.; Wilkinson, W.; Yao, J.; Yi,
  T.~J.; Sandvik, A.~W.; Alvarez, G.; Melko, R.~G.; Carrasquilla, J.; Franz,
  M.; and Amin, M.~H. 2024.
\newblock Computational supremacy in quantum simulation.
\newblock arXiv:2403.00910.

\bibitem[{King et~al.(2021)King, Raymond, Lanting, Isakov, Mohseni,
  Poulin-Lamarre, Ejtemaee, Bernoudy, Ozfidan, Smirnov
  et~al.}]{king2021scaling}
King, A.~D.; Raymond, J.; Lanting, T.; Isakov, S.~V.; Mohseni, M.;
  Poulin-Lamarre, G.; Ejtemaee, S.; Bernoudy, W.; Ozfidan, I.; Smirnov, A.~Y.;
  et~al. 2021.
\newblock Scaling advantage over path-integral Monte Carlo in quantum
  simulation of geometrically frustrated magnets.
\newblock \emph{Nature com.}, 12(1): 1113.

\bibitem[{King et~al.(2022)King, Suzuki, Raymond, Zucca, Lanting, Altomare,
  Berkley, Ejtemaee, Hoskinson, Huang, Ladizinsky, MacDonald, Marsden, Oh,
  Poulin-Lamarre, Reis, Rich, Sato, Whittaker, Yao, Harris, Lidar, Nishimori,
  and Amin}]{King_2022}
King, A.~D.; Suzuki, S.; Raymond, J.; Zucca, A.; Lanting, T.; Altomare, F.;
  Berkley, A.~J.; Ejtemaee, S.; Hoskinson, E.; Huang, S.; Ladizinsky, E.;
  MacDonald, A. J.~R.; Marsden, G.; Oh, T.; Poulin-Lamarre, G.; Reis, M.; Rich,
  C.; Sato, Y.; Whittaker, J.~D.; Yao, J.; Harris, R.; Lidar, D.~A.; Nishimori,
  H.; and Amin, M.~H. 2022.
\newblock Coherent quantum annealing in a programmable
  2,000{\hspace{0.167em}}qubit Ising chain.
\newblock \emph{Nature Physics}, 18(11): 1324--1328.

\bibitem[{Kirkpatrick, Gelatt~Jr, and
  Vecchi(1983)}]{kirkpatrick1983optimization}
Kirkpatrick, S.; Gelatt~Jr, C.~D.; and Vecchi, M.~P. 1983.
\newblock Optimization by simulated annealing.
\newblock \emph{science}, 220(4598): 671--680.

\bibitem[{Li, Chen, and Koltun(2018)}]{Li2018COG}
Li, Z.; Chen, Q.; and Koltun, V. 2018.
\newblock Combinatorial Optimization with Graph Convolutional Networks and
  Guided Tree Search.
\newblock In \emph{Neural Information Processing Systems}.

\bibitem[{Lucas(2014)}]{lucas2014ising}
Lucas, A. 2014.
\newblock Ising formulations of many NP problems.
\newblock \emph{Frontiers in physics}, 5.

\bibitem[{Mugel et~al.(2021)Mugel, Abad, Bermejo, S{\'a}nchez, Lizaso, and
  Or{\'u}s}]{mugel2021hybrid}
Mugel, S.; Abad, M.; Bermejo, M.; S{\'a}nchez, J.; Lizaso, E.; and Or{\'u}s, R.
  2021.
\newblock Hybrid quantum investment optimization with minimal holding period.
\newblock \emph{Scientific Reports}, 11(1): 1--6.

\bibitem[{Oshiyama and Ohzeki(2022)}]{oshiyama2022benchmark}
Oshiyama, H.; and Ohzeki, M. 2022.
\newblock Benchmark of quantum-inspired heuristic solvers for quadratic
  unconstrained binary optimization.
\newblock \emph{Scientific reports}, 12(1): 2146.

\bibitem[{Pan et~al.(2021)Pan, Zhou, Zhou, and Zhang}]{Panfeng2021}
Pan, F.; Zhou, P.; Zhou, H.-J.; and Zhang, P. 2021.
\newblock Solving statistical mechanics on sparse graphs with feedback-set
  variational autoregressive networks.
\newblock \emph{Phys. Rev. E}, 103: 012103.

\bibitem[{Panchenko(2013)}]{panchenko2013sherrington}
Panchenko, D. 2013.
\newblock \emph{The sherrington-kirkpatrick model}.
\newblock Springer Science \& Business Media.

\bibitem[{Patel et~al.(2020)Patel, Chen, Canoza, and
  Salahuddin}]{patel2020ising}
Patel, S.; Chen, L.; Canoza, P.; and Salahuddin, S. 2020.
\newblock Ising model optimization problems on a FPGA accelerated restricted
  Boltzmann machine.
\newblock \emph{arXiv preprint arXiv:2008.04436}.

\bibitem[{Peruzzo et~al.(2014)Peruzzo, McClean, Shadbolt, Yung, Zhou, Love,
  Aspuru-Guzik, and O’brien}]{peruzzo2014variational}
Peruzzo, A.; McClean, J.; Shadbolt, P.; Yung, M.-H.; Zhou, X.-Q.; Love, P.~J.;
  Aspuru-Guzik, A.; and O’brien, J.~L. 2014.
\newblock A variational eigenvalue solver on a photonic quantum processor.
\newblock \emph{Nature communications}, 5(1): 4213.

\bibitem[{Preskill(2018)}]{Preskill_2018}
Preskill, J. 2018.
\newblock Quantum Computing in the {NISQ} era and beyond.
\newblock \emph{Quantum}, 2: 79.

\bibitem[{Rother et~al.(2007)Rother, Kolmogorov, Lempitsky, and
  Szummer}]{rother2007optimizing}
Rother, C.; Kolmogorov, V.; Lempitsky, V.; and Szummer, M. 2007.
\newblock Optimizing binary MRFs via extended roof duality.
\newblock In \emph{2007 IEEE conference on computer vision and pattern
  recognition}, 1--8. IEEE.

\bibitem[{Santoro and Tosatti(2006)}]{Santoro2006}
Santoro, G.~E.; and Tosatti, E. 2006.
\newblock Optimization using quantum mechanics: quantum annealing through
  adiabatic evolution.
\newblock \emph{Journal of Physics A: Mathematical and General}, 39(36): R393.

\bibitem[{Schuetz, Brubaker, and Katzgraber(2021)}]{Schuetz2021PIGNN}
Schuetz, M. J.~A.; Brubaker, J.~K.; and Katzgraber, H.~G. 2021.
\newblock Combinatorial Optimization with Physics-Inspired Graph Neural
  Networks.
\newblock \emph{Nature Machine Intelligence}, 4: 367--377.

\bibitem[{Schuetz et~al.(2022)Schuetz, Brubaker, Zhu, and
  Katzgraber}]{Schuetz2022}
Schuetz, M. J.~A.; Brubaker, J.~K.; Zhu, Z.; and Katzgraber, H.~G. 2022.
\newblock Graph coloring with physics-inspired graph neural networks.
\newblock \emph{Phys. Rev. Res.}, 4: 043131.

\bibitem[{Tabi et~al.(2021)Tabi, Marosits, Kallus, Vaderna, G{\'o}dor, and
  Zimbor{\'a}s}]{tabi2021evaluation}
Tabi, Z.~I.; Marosits, {\'A}.; Kallus, Z.; Vaderna, P.; G{\'o}dor, I.; and
  Zimbor{\'a}s, Z. 2021.
\newblock Evaluation of Quantum Annealer Performance via the Massive MIMO
  Problem.
\newblock \emph{IEEE Access}, 9: 131658--131671.

\bibitem[{Tanaka, Tomiya, and Hashimoto(2023)}]{Akinori2023}
Tanaka, A.; Tomiya, A.; and Hashimoto, K. 2023.
\newblock \emph{Deep Learning and Physics}.
\newblock Springer Singapore.

\bibitem[{Tasseff et~al.(2022)Tasseff, Albash, Morrell, Vuffray, Lokhov, Misra,
  and Coffrin}]{Tasseff2022}
Tasseff, B.; Albash, T.; Morrell, Z.; Vuffray, M.; Lokhov, A.~Y.; Misra, S.;
  and Coffrin, C. 2022.
\newblock {On the Emerging Potential of Quantum Annealing Hardware for
  Combinatorial Optimization}.
\newblock ArXiv:2210.04291.

\bibitem[{Tasseff et~al.(2024)Tasseff, Albash, Morrell, Vuffray, Lokhov, Misra,
  and Coffrin}]{tasseff2024emerging}
Tasseff, B.; Albash, T.; Morrell, Z.; Vuffray, M.; Lokhov, A.~Y.; Misra, S.;
  and Coffrin, C. 2024.
\newblock On the emerging potential of quantum annealing hardware for
  combinatorial optimization.
\newblock \emph{Journal of Heuristics}, 1--34.

\bibitem[{Thai et~al.(2022)Thai, Thai, Vu, and Dinh}]{thai2022fasthare}
Thai, P.; Thai, M.~T.; Vu, T.; and Dinh, T.~N. 2022.
\newblock FastHare: Fast Hamiltonian Reduction for Large-scale Quantum
  Annealing.
\newblock arXiv:2205.05004.

\bibitem[{Tiunov, Ulanov, and Lvovsky(2019)}]{tiunov2019annealing}
Tiunov, E.~S.; Ulanov, A.~E.; and Lvovsky, A. 2019.
\newblock Annealing by simulating the coherent Ising machine.
\newblock \emph{Optics express}, 27(7): 10288--10295.

\bibitem[{Wu, Wang, and Zhang(2019)}]{wu2019solving}
Wu, D.; Wang, L.; and Zhang, P. 2019.
\newblock Solving statistical mechanics using variational autoregressive
  networks.
\newblock \emph{Physical review letters}, 122(8): 080602.

\bibitem[{Yarkoni et~al.(2021)Yarkoni, Alekseyenko, Streif, Von~Dollen,
  Neukart, and B{\"a}ck}]{yarkoni2021multi}
Yarkoni, S.; Alekseyenko, A.; Streif, M.; Von~Dollen, D.; Neukart, F.; and
  B{\"a}ck, T. 2021.
\newblock Multi-car paint shop optimization with quantum annealing.
\newblock In \emph{2021 IEEE International Conference on Quantum Computing and
  Engineering (QCE)}, 35--41. IEEE.

\bibitem[{Zhong et~al.(2020)Zhong, Wang, Deng, Chen, Peng, Luo, Qin, Wu, Ding,
  Hu et~al.}]{zhong2020quantum}
Zhong, H.-S.; Wang, H.; Deng, Y.-H.; Chen, M.-C.; Peng, L.-C.; Luo, Y.-H.; Qin,
  J.; Wu, D.; Ding, X.; Hu, Y.; et~al. 2020.
\newblock Quantum computational advantage using photons.
\newblock \emph{Science}, 370(6523): 1460--1463.

\bibitem[{Zhou and Zhang(2022)}]{zhou2022noise}
Zhou, Y.; and Zhang, P. 2022.
\newblock Noise-Resilient Quantum Machine Learning for Stability Assessment of
  Power Systems.
\newblock \emph{IEEE Transactions on Power Systems}.

\end{thebibliography}
\end{document}